\documentclass{article}
\pdfoutput=1
\usepackage{jcappub}

\def\lbldef#1#2{\expandafter\gdef\csname #1\endcsname {#2}}

\def\href#1#2{#2}

\usepackage{amsmath}
\usepackage{amssymb}
\usepackage{graphicx}% Include figure files
\usepackage{color}

\newcommand{\neff}{{N_{\rm eff}}}

\newcommand{\be}{\begin{equation}}
\newcommand{\ee}{\end{equation}}
\newcommand{\bea}{\begin{eqnarray}}
\newcommand{\eea}{\end{eqnarray}}
\newcommand{\bc}{\begin{center}}
\newcommand{\ec}{\end{center}}

\newcommand{\lik}
  {{\mathcal{L}}}
\newcommand{\ml}
  {{\rm{ML}}}
\newcommand{\rmn}
  {\rm}
\newcommand{\diff}
  {{\rmn{d}}}
\newcommand{\vect}[1]
  {\mbox{\boldmath ${#1}$}}
\newcommand{\pars}
  {\vect{\theta}}
\newcommand{\diracdelta}
  {\delta_{\rm{D}}}
\newcommand{\parsone}
  {\vect{\phi}}
\newcommand{\parstwo}
  {\vect{\psi}}

\newcommand{\bayesfactor} 
  {\frac{E_1}{E_2}}
\newcommand{\prob}{{\rm Pr}}
\newcommand{\data}{\vect{d}}

 \title{(Lack of) Cosmological evidence for dark radiation after Planck} 
\author[a,b]{Licia Verde,}
\author[c]{Stephen M. Feeney,}
\author[d,e]{Daniel J. Mortlock}
\author[c,f]{and Hiranya V. Peiris}
\emailAdd{liciaverde@icc.ub.edu}
\emailAdd{stephen.feeney.09@ucl.ac.uk}
\emailAdd{mortlock@ic.ac.uk}
\emailAdd{h.peiris@ucl.ac.uk}

\affiliation[a]{ICREA \& ICC, Institut de Ciencies del Cosmos, Universitat de Barcelona (IEEC-UB), Marti i Franques 1, Barcelona 08028, Spain}
\affiliation[b]{Theory Group, Physics Department, CERN, CH-1211, Geneva 23, Switzerland}
\affiliation[c]{Department of Physics and Astronomy, University College London, London WC1E 6BT, U.K.}
\affiliation[d]{Astrophysics Group, Imperial College London, Blackett Laboratory, Prince Consort Road, London SW7 2AZ, U.K.}
\affiliation[e]{Department of Mathematics, Imperial College London, London SW7 2AZ, U.K.}
\affiliation[f]{Kavli Institute for Theoretical Physics, Kohn Hall, University of California, Santa Barbara, CA 93106, USA}

\abstract{We use Bayesian model comparison to determine whether extensions to Standard-Model neutrino physics -- primarily additional effective numbers of neutrinos and/or massive neutrinos -- are merited by the latest cosmological data. Given the significant advances in cosmic microwave background (CMB) observations represented by the {\it Planck} data, we examine whether {\it Planck} temperature and CMB lensing data, in combination with lower redshift data, have  strengthened (or weakened) the previous findings. We conclude that  the state-of-the-art cosmological data do not show evidence for deviations from the standard  ($\Lambda$CDM) cosmological model (which has three massless neutrino families). This does not mean  that the model is necessarily correct -- in fact we know it is incomplete as neutrinos are not massless -- but it does imply that deviations from the standard model (e.g., non-zero neutrino mass) are too small compared to the current experimental uncertainties to be inferred from cosmological data alone.}
\begin{document}
\maketitle

%%%%%%%%%%%%%%%%%%%%%%%%%%%%%%%%%%%%%%%%%%%%%%%%%%

\section{Introduction}

Exploring non-standard neutrino properties (including a significantly non-zero mass) requires extending both the concordance cosmological model, $\Lambda$CDM, and the Standard Model of particle physics, and is therefore an issue of model comparison rather than parameter estimation. Model comparison must be performed within the Bayesian framework to be self-consistent~\cite{Cox}, and the relevant quantity to consider in this context is the model-averaged likelihood, referred to here as the Bayesian {\it Evidence}, $E$. Under the assumption of equal {\it a priori} model probabilities, the ratio of Evidence\footnote{We capitalize the Bayesian Evidence to distinguish it from the colloquial ``evidence".} values for two models, given the same data, quantifies the relative odds of these models being the correct description of the observations. 
The use of the Bayesian model comparison framework in cosmology has become popular over the last two decades: e.g., see Refs.~\cite{Jaffe:1996,bridle02, LiddleEvidence04, Parkinson06, Trotta1,Trotta2} and references therein. In previous work~\cite{paper1}, we considered the cosmological Bayesian Evidence for non-standard neutrino properties using pre-{\it Planck} cosmological data. In this paper we update our previous work by considering a compilation of post-{\it Planck} \cite{PlanckPaper12013} cosmological data, which have significantly extended the state-of-the-art compared with the data used in our previous analysis.

The Standard Model of particle physics has three massless neutrinos. Beyond-Standard-Model physics (or uncomfortable fine-tuning) is needed to give neutrinos a non-zero mass, and extensions of the Standard Model include the possibility of more than three neutrino species. The standard cosmological model, $\Lambda$CDM, also has three massless neutrino families, but quantifies their effects through the effective number of species, $N_{\rm eff}=3.046$. This differs from the number of neutrino species, $N_{\nu}=3$, to account for QED effects, for neutrinos being not completely decoupled during electron-positron annihilation  and other small effects (see, e.g., Refs.~\cite{lesgourguespastor,mangano}).
For cosmological observations, a deviation from the Standard Model prediction for $N_{\rm eff}$  does not necessarily imply new neutrino physics: e.g., any non-standard early-Universe expansion history due to non-standard energy-density can parametrized in terms of $N_{\rm eff}$. This ``dark radiation" has been the subject of great interest in the pre-{\it Planck} years: for a review, see Ref.~\cite{Abazajian:2012ys} and references therein, and the discussion and references in Ref.~\cite{paper1}. 

The pre-{\it Planck}  cosmological data  have been often interpreted as supporting the case for dark radiation, as well as one or more sterile neutrinos. Post-{\it Planck}, the data leave less freedom to the sterile-neutrino interpretation~\cite{planckparameterspaper}, but Bayesian parameter estimates of models with $N_{\rm eff}$ as a free parameter still leave room for dark radiation (see, e.g., Refs.~\cite{weinbergneff, Said:2013hta, DiValentino:2013qma, Kelso:2013paa, Mastache:2013iha, DiBari:2013dna, Archidiacono:2013fha}). The official {\it Planck} analysis (and other work considering {\it Planck} data) concentrated on parameter estimation and did not address the model comparison issue. It is therefore important to consider the newly-available data in the framework of model comparison, and examine how these findings might depend on the assumed priors.
In addition we note that, while in  the standard cosmological model the neutrinos have traditionally been assumed to be  massless, the {\it Planck} collaboration chose their baseline, standard $\Lambda$CDM model to have a non-zero total neutrino mass: $M_{\nu}=0.06$ eV~\cite{planckparameterspaper},  which is close to the minimum value allowed by neutrino-oscillation experiments (see, e.g., Ref.~\cite{PDG} and references therein). Because of the significant increase in the precision of recent cosmological data, this choice implies small shifts in the best-fit cosmological parameters compared to a model in which $M_{\nu}=0$. It is interesting to consider whether current cosmological data exhibit any evidence for this ``paradigm shift" (though this choice is well-motivated by non-cosmological data).

This paper is organized as follows. In Sec.~\ref{sec:methods} we describe the methods used in our analysis, their interpretation, and the datasets we consider. In Sec.~\ref{sec:results} we present our results, and in Sec.~\ref{sec:futureconclusions} we present our conclusions and discuss future avenues to further constrain neutrino properties in the context of cosmological model comparison.

%%%%%%%%%%%%%%%%%%%%%%%%%%%%%%%%%%%%%%%%%%%%%%%%%%

\section{Methods and datasets}
\label{sec:methods}
We wish to answer the following question: do the latest cosmological data require the inclusion of extra parameters (describing non-standard neutrino properties) beyond the simple $\Lambda$CDM model? Rather than being an issue of parameter estimation -- i.e., the determination of the most probable values for the extra parameter(s) within the context of a single model -- this is a question of model comparison.

Within a Bayesian framework the key model comparison quantity is the Bayes factor, which is the ratio of the Evidence values for two different models (see Sec.~\ref{sec:evidence_ratio} below). In general, the Evidence is the result of a multi-dimensional integral over the model parameters, $\pars$, the evaluation of which can be computationally expensive, and specialized algorithms have been developed to compute it efficiently~\cite{nestedsampling1,nestedsampling2}. Here, however, the Bayes factors can be computed much more efficiently because the models we are comparing have some parameters in common, and indeed are nested (see Sec.~\ref{sec:SDDR} below).  The parameter space $\pars = (\parsone, \parstwo)$ is partitioned into the common parameters $\parsone$ (in this case, those of vanilla $\Lambda$CDM: the physical densities of baryons and cold dark matter, $\Omega_b h^2$ and $\Omega_c h^2$; the angular scale of the sound horizon at last-scattering, $\Theta_A$; the integrated optical depth to last-scattering, $\tau$; and the matter power spectrum spectral slope and amplitude, $n_s$ and $A_s$) and the extra parameters $\parstwo$ (in this case, those describing the additional neutrino properties, $N_{\rm eff}$, $M_{\nu}$, etc.).  The question asked above is addressed by determining whether the data support a model, $M_2$, in which the extra parameters are not restricted to their fiducial $\Lambda$CDM values.

As with all Bayesian methods, the Bayes factor between these two models depends on the prior on the model parameters. While in some cases the priors can be physically motivated, this is not always possible; in the latter case, the Bayesian answer to the model comparison question calculated with specific prior choices cannot be considered definitive. We will therefore also consider a statistic -- the profile likelihood ratio (PLR) -- which depends only on the likelihood, and thus is prior-independent. While the PLR does not provide a self-consistent model-selection paradigm, it is useful for assessing whether the Bayesian results are driven by the prior choice or the data.

%%%%%%%%%%%%%%%%%%%%%%%%%%%%%%%%%%%%%%%%%%%%%%%%%%

\subsection{Bayes factor}
\label{sec:evidence_ratio}
The probability distribution ($\prob$) for a set of parameters, $\pars$, given a model, $M$, and data, $\data$, is the posterior, $P=\prob(\pars|\data,M)$. Bayes' theorem relates the posterior to the likelihood, ${\cal L} \equiv \prob(\data|\pars,M)$, via the prior, $\pi \equiv \prob(\pars|M)$:
\begin{equation}
\prob(\pars|\data,M) = \frac{\prob(\pars|M) \, \prob(\data|\pars,M)}{\prob(\data|M)}\,.
\end{equation}
The Bayesian Evidence or model-averaged likelihood, $E\equiv \prob(\data|M)$, normalizes the parameter posterior, and is given by
\begin{equation}
E = \int \prob(\pars|M) \,\prob(\data|\pars,M)\mathrm{d}\pars\,.
\label{eq:evidence}
\end{equation}
Using this expression we can write the ratio of probabilities of two models in the light of the data as
\begin{equation}
\frac{\prob(M_1|\data)}{\prob(M_2|\data)} = \frac{\prob(M_1)}{\prob(M_2)}\frac{\prob(\data|M_1)}{\prob(\data|M_2)} = \frac{\prob(M_1)}{\prob(M_2)} \frac{E_1}{E_2}\,,
\end{equation}
where the {\em a priori} probability ratio of the two models, $\prob(M_1)/\prob(M_2)$, is typically set to unity, and $\prob(\data|M_1)/\prob(\data|M_2) = E_1 / E_2$ is the Bayes factor between the two models. The Bayes factor, or ratio of the Evidence values for two models given the same data, therefore expresses the relative odds that these models are responsible for the observed state of the Universe.

It is important to note here that Bayesian model comparison does not automatically reward {\em simpler} models, but rather rewards {\em predictive} models (see, e.g., Ref.~\cite{Liddle_etal:2007}). 
If an extension to $\Lambda$CDM predicts observables (e.g., CMB power spectra) that are indistinguishable from those of $\Lambda$CDM (and the two models' likelihoods are hence equal), then the Evidence values for the two models are equal. 
More generally, consider a simple model, $M_1$, and an extended model, $M_2$.
If the extra parameters in the extended model have no influence on the likelihood, so that $\prob(\data | \parsone, \parstwo, M_2) = \prob(\data |\parsone, M_1)$, and the marginalized prior for the common parameters in the extended model, $\prob(\parsone | M_2) = \int \prob(\parsone, \parstwo |M_2) \, \diff \parstwo$, is the same as their prior in the simpler model, $\prob(\parsone | M_1)$, then the Evidence for the extended model is given by
\begin{eqnarray*}
E_2 & = & \int \int \prob(\parsone , \parstwo | M_2) \, \prob(\data| \parsone, \parstwo, M_2) \, {\rm d}\parsone \, {\rm d}\parstwo \\
& = & \int \left[ \int \prob(\parsone, \parstwo | M_2) \, {\rm d}\parstwo \right] \prob(\data| \parsone, M_1) \,  {\rm d}\parsone \\
& = & \int \prob(\parsone | M_1) \, \prob(\data| \parsone, M_1) \, {\rm d}\parsone\\
& = & E_1.
\end{eqnarray*}
Within the context of Bayesian model comparison a model does not have an intrinsic complexity; it is only the ability of a model to predict the results of measurements that is important. If the available observations are equally likely within the context of two models then neither is preferred by the data, regardless of their internal structure.

In situations where the evidence values differ
we use a slightly modified Jeffreys'
scale~\cite{Jeffrey,kassraftery} to interpret our Evidence ratios. This scale, which classifies Evidence ratios from ``not worth a bare mention'' to ``highly significant'', is defined in Table~\ref{tab:Jeffrey}).
Care should be taken when using thresholded scales such as the Jeffreys' scale, which introduce sharp decision-making boundaries. It is used here only as a loose classification of the strength of the Evidence ratios, as we are comparing a large number of model and dataset combinations.

\begin{table}[tb]
\centering
 \begin{tabular}{ccc}
 \hline
$ | \ln (E_1/E_2)|$ & interpretation & betting odds\\
 \hline
 $<1$ & not worth a bare mention &$<3:1$ \\
 $1-2.5$ & substantial & $\sim 3:1$\\
 $2.5-5$ & strong &$>12:1$ \\
 $>5$ & highly significant & $>150:1$\\
  \hline
\end{tabular}
 \caption{The slightly modified Jeffreys' scale we use for interpreting the Evidence ratios.}
 \label{tab:Jeffrey}
\end{table}

%%%%%%%%%%%%%%%%%%%%%%%%%%%%%%%%%%%%%%%%%%%%%%%%%%

\subsection{Savage-Dickey Density Ratio}
\label{sec:SDDR}

The extensions to $\Lambda$CDM considered in this work are clearly related to $\Lambda$CDM itself: $\Lambda$CDM is a simpler version of the extended models, and is in fact ``nested'' within the extensions.  A simpler model $M_1$ can be considered to be nested within an extended model $M_2$ if:

\begin{itemize}
    \item the parameters which describe model $M_2$ are separable into two disjoint subsets: the base parameters, $\parsone$, fully describing $M_1$, and some extra parameters, $\parstwo$, unique to $M_2$;
    \item there exists a combination of the extra parameters, $\parstwo = \parstwo_1$, for which the second model reduces identically to the simpler model so that the likelihood $\prob(\data|\parsone , M_1)=\prob(\data|\parsone,\parstwo_1, M_2)$ for all $\data$;
    \item the prior distributions of the extra parameters in $M_2$ are separable from the prior distributions of the base parameters: $\prob(\parsone,\parstwo | M_2)=\prob(\parsone | M_2)\prob(\parstwo | M_2)$, with $\prob(\parstwo_1 | M_2) > 0$;
    \item the prior distributions of the common parameters are the same: $\prob(\parsone | M_1)=\prob(\parsone | M_2)$.
\end{itemize}

In the case of nested models, the Bayes factor is
\begin{eqnarray}
\label{equation:bf_sddr}
\bayesfactor = \frac{\prob(\data | M_1)}{\prob(\data | M_2)}
 & = & \frac{
   \int \prob(\parsone^{\prime} | M_1) \, 
    \prob(\data | \parsone^{\prime}, M_1) 
    \, \diff \parsone^{\prime}
  }
  {
    \int \prob(\parsone^{\prime\prime}, \parstwo^{\prime\prime} | M_2) \, 
    \prob(\data | \parsone^{\prime\prime}, \parstwo^{\prime\prime}, M_2) \, 
    \diff \parsone^{\prime\prime} \, \diff \parstwo^{\prime\prime}
  }\\
  &=& \frac{
   \int \prob(\parsone^{\prime} | M_1) \, 
   \prob(\data | \parsone^{\prime}, \parstwo_1,M_2) 
    \, \diff \parsone^{\prime}
  }
  {
    \int \prob(\parsone^{\prime\prime} | M_1) 
      \prob( \parstwo^{\prime\prime} | M_2) \, 
    \prob(\data | \parsone^{\prime\prime}, \parstwo^{\prime\prime},M_2) 
    \, \diff \parsone^{\prime\prime} \, \diff \parstwo^{\prime\prime}
  } \, ,
\end{eqnarray}
where  in the second line we have used the fact that the models are nested, and care has been taken to avoid confusing the distinct dummy integration variables in the numerator and the denominator. Multiplying both numerator and denominator by
$\prob(\parstwo_1 | M_2)$ then gives
\begin{equation}
\bayesfactor  =  \frac{1}{\prob( \parstwo_1 | M_2)} 
  \frac{
   \int \prob(\parsone^{\prime} | M_1) \, \prob( \parstwo_1 | M_2) \,
    \prob(\data | \parsone^{\prime}, \parstwo_1,M_2) \, 
    \diff \parsone^{\prime}
  }
  { 
    \int \prob(\parsone^{\prime\prime} | M_1) 
      \, \prob( \parstwo^{\prime\prime} | M_2) \, 
    \prob(\data| \parsone^{\prime\prime}, \parstwo^{\prime\prime},M_2) 
    \, \diff \parsone^{\prime\prime} \, \diff \parstwo^{\prime\prime}
  } \, .
\end{equation}  
By comparison, the normalized parameter posterior for $(\parsone, \parstwo)$ under the second model is 
\begin{equation}
\prob(\parsone, \parstwo | \data, M_2)
  = \frac{\prob(\parsone | M_1) \, \prob( \parstwo | M_2) \,
    \prob(\data | \parsone, \parstwo_1,M_2)}
  { \int \prob(\parsone^{\prime\prime} | M_1) 
     \, \prob( \parstwo^{\prime\prime} | M_2) \, 
    \prob(\data| \parsone^{\prime\prime}, \parstwo^{\prime\prime},M_2) 
     \, \diff \parsone^{\prime\prime} \, \diff \parstwo^{\prime\prime}}
\end{equation}
for all values of $\parstwo$ (and, in particular, for $\parstwo = \parstwo_1$). Hence 
\begin{equation}
\bayesfactor 
=\frac{
   \int \prob(\parsone^\prime, \parstwo_1 | \data, M_2) \, 
    \diff \parsone^\prime 
  }
  { \prob( \parstwo_1 | M_2) }\, .
\end{equation}
But integrating a posterior distribution over some sub-set of the parameters is simply marginalization, and in this case the integral in the numerator reduces to the marginalized posterior $\prob(\parstwo_1 | \data, M_2)$.
Hence the evidence ratio between two nested models is given by 
\begin{equation}
\label{equation:sddr}
\bayesfactor
  = \frac{\prob(\parstwo_1 | \data, M_2)}{\prob(\parstwo_1 | M_2)}
  = \left.
    \frac{\prob(\parstwo | \data, M_2)}{\prob(\parstwo | M_2)}
    \right|_{\parstwo = \parstwo_1}\, .
\end{equation}
This is the Savage--Dickey density ratio (SDDR; Ref.~\cite{Dickey1971}\footnote{Dickey attributed this result to Savage, hence the name.  The implications of this result are further explored in e.g., Refs.~\cite{VerdinelliWasserman95, OHagan, MarinRobert10, Trotta2}.}).  
Using the SDDR makes it possible to perform rigorous model comparison between nested models without the need for evaluating numerically-intensive multi-dimensional integrals over the two models' parameters.

It may seem surprising that such an apparently complicated ratio of integrals has such a simple expression, but the interpretation of this final result is  clear. The simpler nested model is preferred if, within the context of the more-complicated model, the relevant data result in an increased probability that $\parstwo = \parstwo_1$; the more complicated model is preferred if  the data imply that $\parstwo = \parstwo_1$ is disfavoured compared with other values for the extra parameters in the second model.  Note also that this gives an alternative derivation of the result given above for the case in which the extra parameters have no influence on the likelihood; in that case the distributions $\prob(\parstwo | \data, M_2)$ and $\prob(\parstwo | M_2)$ must be the same, again yielding $E_2 = E_1$.

Critically, the SDDR given in Eq.~\ref{equation:sddr} can be estimated from samples drawn from the posterior distribution $\prob(\pars | \data, M_2)$, such as a Markov Chain Monte Carlo (MCMC) outputs. For detailed and high-precision implementations see, e.g., Refs.~\cite{vanHaasteren:2009yg, WeinbergMD}. We will use a simpler implementation which still yields very accurate results in the relevant setting, as discussed in Sec.~\ref{sec:resultsevidence}.

%%%%%%%%%%%%%%%%%%%%%%%%%%%%%%%%%%%%%%%%%%%%%%%%%%

\subsection{Profile likelihood ratio} 
\label{sec:PLR}

In the absence of well-motivated parameter priors, we can use the likelihood alone to assess the significance of any deviations of the extra parameters ($\parstwo$) from their fiducial values ($\parstwo_1$).  This can be seen by considering an extreme alternative model in which the extra parameters are fixed at some value $\parstwo_2$, implying a parameter prior of the form $\prob(\parsone, \parstwo | M_2) = \prob(\parsone | M_2) \, \diracdelta(\parstwo - \parstwo_2)$.  The Evidence for this model immediately simplifies to $E_2 = \int \prob(\parsone^\prime | M_2) \, \prob(\data | \parsone^\prime, \parstwo_2) \, \diff \parsone^\prime$, and would be maximized if $\prob(\parsone | M_2) = \diracdelta[\parsone - \parsone_\ml(\parstwo_2, \data)]$, where $\parsone_\ml(\parstwo_2, \data)$ is the maximum likelihood (ML) value of $\parsone$ under the restriction that $\parstwo = \parstwo_2$.  While it would be unreasonable to assess a model by tuning it to the same data being used to test it, this does provide an upper bound on the Evidence, as $E_2 \leq \prob[\data | \parsone_\ml(\parstwo_2, \data), \parstwo_2 ] = \lik_{\rm max} (\parstwo_2)$.  

It is now possible to examine how $\lik_{\rm max} (\parstwo_2)$ depends on the value of $\parstwo_2$.  The maximum possible value of $\lik_{\rm max} (\parstwo_2)$ occurs if $\parstwo_2 = \parstwo_\ml(\data)$, the overall maximum-likelihood model, in which case $\lik_{\rm max} (\parstwo_2) = \prob[\data | \parsone_\ml(\data), \parstwo_\ml(\data)] = \lik_{\rm max}$.  The normalized ratio
\begin{equation}
\frac{ \prob[\data | \parsone_\ml(\parstwo_2, \data), \parstwo_2 ] }{ \prob[\data | \parsone_\ml(\data), \parstwo_\ml(\data)] } = \frac{ \lik_{\rm max} (\parstwo_2) }{ \lik_{\rm max} }
\end{equation}
then gives a heuristic assessment of how plausible a model with $\parstwo = \parstwo_2$ is given the data.  This is known as the profile likelihood ratio (PLR; \cite{PLR}).

The PLR has an interpretation similar to the $\Delta \chi^2$ for nested models, where the effective chi-square is identified with $-2 \ln \prob(\data|\parsone,\parstwo,M_2)$ and, by construction, is prior-independent. Assuming Gaussian statistics one may associate ratios of 0.5 and 2.0 to one- and two-$\sigma$ confidence regions. If, e.g., the PLR at a non-standard value of $\neff$ were two greater than the PLR at $\neff = 3.046$, one could claim $\sim 2\sigma$ evidence in favor of the $\Lambda$CDM+$\neff$ model. It must be noted that these confidence intervals may not have strict frequentist coverage, especially if the likelihood is far from Gaussian; nevertheless, this quantity allows us to assess whether the Evidence results are driven by the data or by the prior.
 
 In practice, the PLR is computed from the output of Monte Carlo Markov chains with the approximation that the conditional maximum likelihood is computed in bins for the interesting parameter (see, e.g., Ref.~\cite{Reid:2009nq,alma,paper1}.)  We verify that the dependence on the bin size is small.

%%%%%%%%%%%%%%%%%%%%%%%%%%%%%%%%%%%%%%%%%%%%%%%%%%

\subsection{Data}

The {\it Planck} satellite mission  \cite{PlanckPaper12013} recently released its temperature data from 2.6 surveys of the sky (15 months). Along with the data, the {\it Planck} Collaboration also released the MCMCs used to sample from the space of possible cosmological parameters  and generate estimates of the posterior mean of each parameter of interest, along with a confidence interval. Here, we use the publicly-available outputs of the {\it Planck} Collaboration's MCMCs.

Several combinations of the {\it Planck} data were analyzed by the Collaboration, and a large number of MCMCs are therefore available. The minimum dataset uses only {\it Planck} temperature data ({\it Planck-lowl}, here shortened to {\it Planck}) in the multipole range $\ell <2500$. Since there are no {\it Planck} polarization data in this first cosmological data release to constrain the optical depth to reionization, $\tau$, WMAP polarization data~\cite{Bennett:2012fp} ({\it lowLike} or {\it WP}) at low $\ell$ are also included; in some cases a $\tau$ prior ({\it tauprior}) is used instead, but we do not use it here. The {\it Planck} data analysis has also produced a reconstruction of the CMB lensing potential through the measurement of the temperature-anisotropy four-point function. The lensing potential probes the amplitude of the large-scale cosmological structure integrated all the way to the present time, thus helping break cosmological degeneracies that are intrinsic to the primary CMB anisotropies.  When the information from the lensing potential is included it is referred to as {\it lensing}. Note that the effects of lensing on the primary anisotropies (smoothing of the higher-order acoustic peaks of the temperature power spectrum and the change in the damping tail) are always included. 

Ground-based telescopes (the Atacama Cosmology Telescope, ACT \cite{ACT}, and the South Pole Telescope, SPT \cite{SPT}) have mapped the sky over small areas at CMB frequencies. These experiments have mapped the CMB damping tail but, more importantly in the context of the {\it Planck} data analysis, have mapped the foregrounds with higher resolution and lower noise than {\it Planck} (albeit not over the full sky).  These data (named {\it highL}) can therefore be used with {\it Planck} data to better constrain the foreground-model parameters and thus reduce degeneracies these parameters may have with the cosmology. Other lower redshift observations can also be used to break cosmological degeneracies. The full set of such data sources used in the {\it Planck} analysis is summarized in Table~\ref{tab:data}. 
 \begin{table}[tb]
\centering
 \begin{tabular}{lcc}
 \hline
 data & shorthand & reference\\
 \hline
 {\it Planck} temperature & Planck & \cite{Plancktemperaturecl}\\
 WMAP low $\ell$ polarization  & lowLike or WP &\cite{Bennett:2012fp}\\
 Prior on optical depth & tauprior & \cite{Hinshaw:2012aka,Plancktemperaturecl}\\
 {\it Planck} reconstructed lensing potential & lensing & \cite{Plancklensing}\\
 ACT and SPT& highL& \cite{ACT, SPT}\\
 Baryon acoustic oscillations compilation& BAO& \cite{Percival:2010, Padmanabhan:2012hf, Blake:2011en,Anderson:2012sa,Beutler:2011hx}\\
 Hubble constant & HST &\cite{Riess:2011yx}\\
 Supernovae &SNLS & \cite{Conley:2011ku}\\
 Supernovae &Union2& \cite{Suzuki:2011hu}\\
 \hline
\end{tabular}
 \caption{Summary of the combinations of datasets, their abbreviation and relevant references.}
 \label{tab:data}
\end{table}

In this paper our ``Gold" set will be {\it Planck} temperature data combined with WMAP polarization information at low $\ell$. We will also report results for other combinations (where available), but we prefer to highlight the Gold dataset for the following reasons.
Firstly, it involves only high-quality CMB observations whose interpretation is based on well-understood, linear physics.
Secondly, the combination of {\it Planck} temperature data and WMAP polarization data (Planck+WP) with {\it highL} data predicts a value for the lensing amplitude which is about 2$\sigma$ higher than the value measured from the convergence power spectrum~\cite{planckparameterspaper}. As the origin of this tension is not yet fully understood, we prefer not to include {\it lensing} and {\it highL} in our Gold dataset. There has also been significant discussion on the fact that the value of the Hubble constant extrapolated from {\it Planck} is in tension with the  local measurement \cite{planckparameterspaper}. We prefer not to combine datasets that are slightly in tension in our Gold set.
 
Other datasets considered by the {\it Planck} Collaboration in combination with CMB data include measurements of the baryon acoustic oscillation feature (BAO) \cite{Percival:2010, Padmanabhan:2012hf, Blake:2011en,Anderson:2012sa,Beutler:2011hx} and two sets of supernovae data, SNLS~\cite{Conley:2011ku} and Union2~\cite{Suzuki:2011hu}. While BAO data are in good agreement with {\it Planck}, the two supernovae datasets give slightly different results when combined with CMB data. For this reason, the two supernovae datasets are always considered separately.

Although we focus on the Gold set, especially in the discussion of the results and in figures,  we report full results for a range of data combinations beyond  the Gold set.

%%%%%%%%%%%%%%%%%%%%%%%%%%%%%%%%%%%%%%%%%%%%%%%%%%

\section{Results}
 \label{sec:results}
We consider the following extensions to the $\Lambda$CDM model: 

\begin{enumerate}
\item the addition of a non-standard effective number of neutrino species,  $\Lambda$CDM+$\neff$ (within which $\Lambda$CDM is nested at $N_{\rm eff} = 3.046$);
\item the addition of a non-standard neutrino number and a primordial Helium abundance  ($Y_P$), not fixed by Big Bang nucleosynthesis, $\Lambda$CDM+$N_{\rm eff} +Y_P$ (within which $\Lambda$CDM is nested at $N_{\rm eff} = 3.046$, $Y_P = 0.24$);
\item a model with three massive neutrinos, $\Lambda$CDM+$M_{\nu}$, where $M_{\nu}$ is the sum of the individual neutrino masses (within which $\Lambda$CDM is nested at $M_\nu = 0$,  or $0.06 $ eV following Ref.~\cite{planckparameterspaper});
\item the addition of both neutrino mass and a non-standard effective number of neutrinos, $\Lambda$CDM+$M_{\nu}$+$N_{\rm eff}$ (within which $\Lambda$CDM is nested at $N_{\rm eff} = 3.046$, $M_\nu = 0$ or 0.06); and
\item a model with one massive sterile neutrino in addition to the three active neutrinos (two massless and one massive), $\Lambda$CDM+$\neff$+$m_{\nu\, {\rm sterile}}^{\rm eff}$ (within which $\Lambda$CDM is nested at $N_{\rm eff} = 3.046$, $m_{\nu\, {\rm sterile}}^{\rm eff} = 0$) -- the two parameters describe a sterile neutrino with freely-varying mass and temperature.
\end{enumerate}

The priors used are those of Ref.~\cite{planckparameterspaper}, which are reported in their Table 1; we summarize the priors on the cosmologically-interesting parameters here in Table~\ref{tab:priors}.

\begin{table}[tb]
\centering
 \begin{tabular}{|l|cccccc|cccc|}
 \hline
   & \multicolumn{6}{c|}{$\parsone$} & \multicolumn{4}{c|}{$\parstwo$} \\
 \hline
parameter& $\Omega_b h^2$&$\Omega_c h^2$&$\Theta_A$&$\tau$& $n_s$&$\ln(10^{10} A_s)$&$M_{\nu}$&$N_{\rm eff}$&$Y_P$&$m_{\nu\,{\rm sterile}}^{\rm eff}$\\
\hline
prior minimum &$0.005$&$0.001$&$0.005$&$0.01$&$0.9$&$2.7$&$0$&$0.05$&$0.1$&$0$\\
prior maximum &$0.1$&$0.99$&$0.1$&$0.8$&$1.1$&$4$&$5$&$10$&$0.5$&$3$\\
\hline
\end{tabular}
 \caption{ Prior ranges used for the cosmologically relevant parameters. The  priors employed are  those of Ref.~\cite{planckparameterspaper}; see their Table 1 for more details.}
 \label{tab:priors}
\end{table}

%%%%%%%%%%%%%%%%%%%%%%%%%%%%%%%%%%%%%%%%%%%%%%%%%%

\subsection{Evidence}
\label{sec:resultsevidence}
As motivated in Sec.~\ref{sec:SDDR} we estimate the Evidence with the SDDR.  The numerical implementation of this estimate, when evaluated from MCMCs, introduces two  approximations: (1) the MCMC is a finite sampling of the posterior, and (2) the conditional maximum of the posterior is calculated for binned values of the parameter of interest. We quantify the error on the Evidence ratio introduced by each of these effects. The MCMC sampling effect can be estimated in a jackknife-like fashion by  evaluating the scatter on the evidence computed from sub-sections of the MCMCs. The binning effect is evaluated by changing the bin size.  We find that both these effects contribute errors of $\sim \pm 0.02$ to the log-evidence ratio for most cases. Errors can reach $\pm 0.1$ for thinned, post-processed (importance-sampled) chains, but even these errors are small compared to the relevant  $\ln (E_1 / E_2)$.

The MCMCs provide very precise estimates of the Evidence ratio, despite not being designed with Evidence calculation in mind. This is a consequence of our studying nested models, and specifically models nested at parameter values fairly close to the peak in the posterior. In these circumstances, the MCMCs characterize the posterior at the nested value very well, providing an accurate and precise estimate of the SDDR. Should the nested value be far from the posterior peak we would expect both the accuracy and precision of the SDDR to suffer, although arguably in this case the model comparison results would be definitively in favor of the more-complex model.

Further confidence in the performance of the SDDR applied to MCMCs can be obtained by direct comparison to the Evidence as calculated by nested sampling. In previous work~\cite{paper1}, we used MultiNest~\cite{Feroz:2007kg,Feroz:2008xx} to calculate the Evidence for various models with extended neutrino physics in the light of CMB observations by WMAP~\cite{Larson:2010gs} and SPT~\cite{Keisler:2011aw}. Both of these collaborations have performed parameter-estimation for the same (or similar\footnote{The prior on $\neff$ employed by WMAP extends to slightly lower values than that used by SPT and the previous work; however, the likelihood is very low in this region of parameter space ($0 < \neff < 1$), and this region can therefore be excluded from the SDDR calculation without significantly changing the results.}) models, and have released the MCMCs from which their parameter constraints are obtained. Calculating the SDDR using these chains, we find that the typical difference between the log-Evidence ratios ($\ln[E_{\Lambda {\rm CDM+}} / E_{\Lambda {\rm CDM+}\neff}]$) obtained via the SDDR and nested sampling is 0.1--0.3, in line with the estimated error in the MultiNest calculation of $\sim 0.3$.

 \begin{table}[tb]
\centering
 \begin{tabular}{|lc|cccccc|}
 \hline
   & & \multicolumn{6}{c|}{evidence ratio: $\ln (E_{\Lambda CDM}/E_{\rm extension})$} \\
 \hline
model extension&data & base&+BAO&+$H_0$&+SNLS&+Union2 &+lensing\\
 \hline
  & Planck &2.12&3.25 P&3.73 P&3.42 P &2.97 P&0.38 P\\
$\Lambda$CDM+$M_{\nu}$&Planck+WP   &2.65 &3.56 P &3.88 P &3.62 P & 3.25 P&\\
(vs $M_{\nu}=0$ eV) &+ lensing & 1.88 & 3.27 P & 3.83 P & 3.48 P & &\\
 & Planck+WP+highL &3.00&3.63 P&3.93 P&3.70 P&3.39 P&\\
 \hline
   & Planck &2.01&3.17 P&3.40 P&3.20 P &2.85 P&0.33 P\\
$\Lambda$CDM+$M_{\nu}$&Planck+WP   &2.53 &3.35 P &3.46 P &3.32 P & 3.06 P&\\
(vs $M_{\nu}=0.06$ eV) &+ lensing & 1.83 & 3.24 P & 3.39 P & 3.20 P &  &\\
 & Planck+WP+highL &2.90&3.43 P&3.51 P&3.40 P&3.22 P&\\
 \hline
  & Planck &-0.43&1.78 P&-1.44 P&-1.89 P& -0.21 P&1.27 P\\
$\Lambda$CDM+$N_{\rm eff}$&Planck+WP  &1.62&1.90&-0.81&1.49&0.96&1.93 P\\
 &+ lensing P &1.93 & 2.36&-0.02&1.91& 1.60&\\
  & Planck+WP+highL &2.00&2.28&0.14 P&1.52 P&1.94 P&2.17\\
  \hline
   & Planck+WP &1.76&2.09 P&1.10 P&&&1.72 P\\
$\Lambda$CDM+$N_{\rm eff}$+$Y_p$&Planck+WP+highL &1.75&2.11 P&0.48 P&&&1.78 P\\
  \hline
   & Planck+WP &4.36&5.11 P&2.80 P&&&3.87 P\\
$\Lambda$CDM+$N_{\rm eff}$+$M_{\nu}$&Planck+WP+highL &5.05&5.59 P&3.73 P&&&4.68 P\\
\hline
$\Lambda$CDM+$\neff$+$m_{\nu\,{\rm sterile}}^{\rm eff}$&Planck+WP+highL &3.52&3.72 P&2.29 P& &&3.38 P\\
  \hline
\end{tabular}
 \caption{Evidence results for all of the model-dataset combinations we consider. A ``P" next to the number or dataset indicates that the results have been obtained by importance sampling (i.e., post-processing) the MCMCs. Note that the extra datasets (BAO, supernovae, etc.) are always added one-by-one to the base combination.  Positive numbers mean that the simpler model is preferred, negative numbers that the more-complicated model is preferred. The errors on the evidence ratios are in most cases about $\pm 0.02$, reaching $\pm 0.1$ for thinned, post-processed chains.}
 \label{tab:fullresultevidence}
\end{table}

\begin{figure}[tb]
\centering
\includegraphics[width=16.5cm]{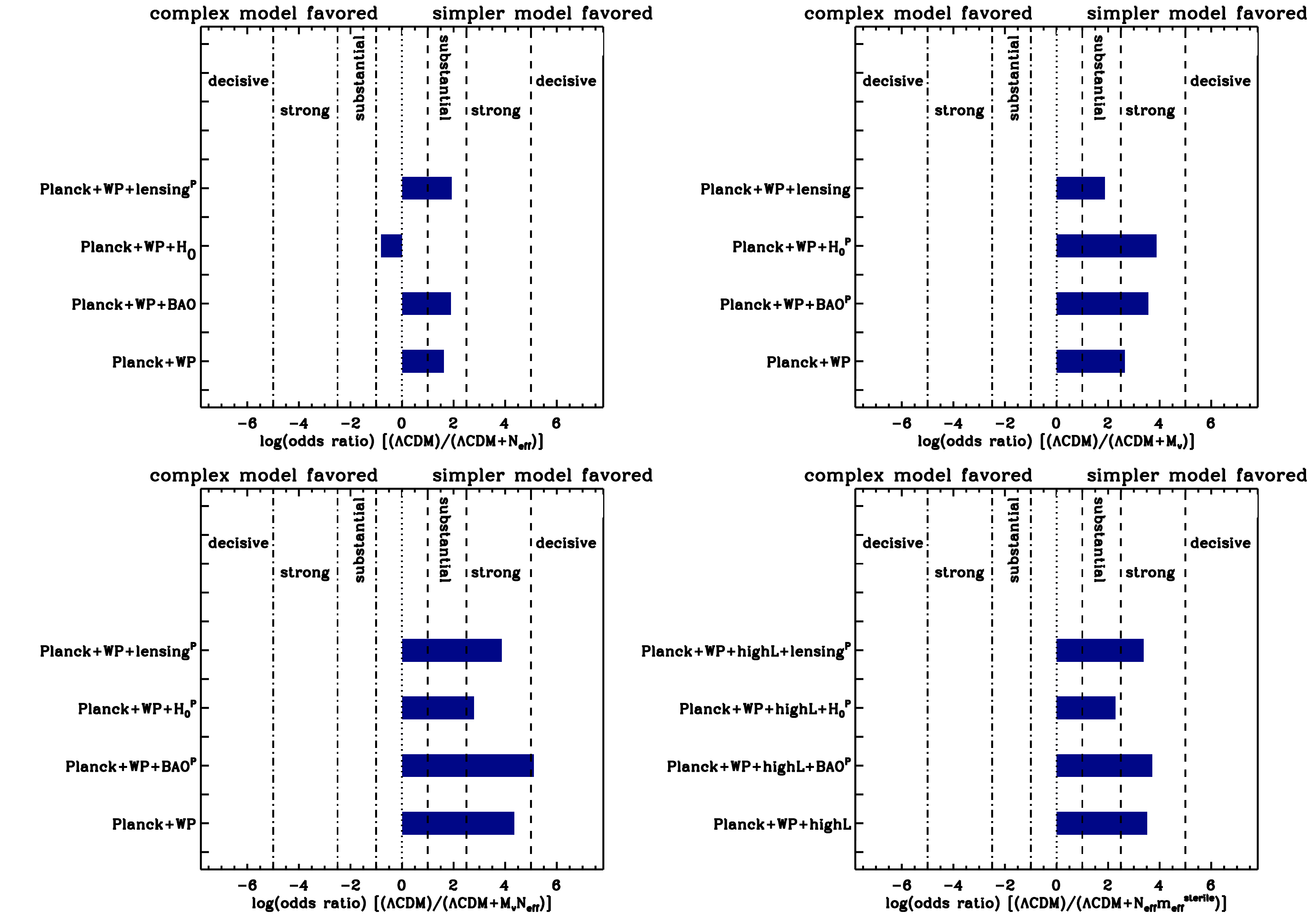}
\caption{Summary of the Evidence ratios for the models discussed in the text for selected dataset combinations. The Evidence for the simplest model ($\Lambda$CDM) is substantial to strong, except when the ``discrepant" $H_0$ measurement is added. In this case there is still no evidence to favor the more-complex $\Lambda$CDM+$N_{\rm eff}$ model.  }
\label{fig:evidencemulti}
\end{figure}

The complete set of Evidence ratios comparing the base $\Lambda$CDM model with the more-complex models is reported in Table~\ref{tab:fullresultevidence}. Fig.~\ref{fig:evidencemulti} also shows a visual comparison of a selection of the table entries. Concentrating first on the $\Lambda$CDM+$\neff$ model comparisons, we see that the Evidence in favor of the simplest model ($\Lambda$CDM) is ``substantial"   to ``strong", except when the $H_0$ measurement is added. The high local measurements of $H_0$ and the low extrapolations from {\it Planck} are known to be discrepant at the $\sim 2 \sigma$ level, a tension that can be relieved by increasing the effective number of neutrino species (and hence the expansion rate at both early and late times~\cite{Hou:2011ec}). Nevertheless, even when considering datasets known to be in tension, there is still no substantial Evidence in favor the more complex $\Lambda$CDM+$N_{\rm eff}$ model. As anticipated in Ref.~\cite{paper1}, the CMB lensing signal offers a powerful  constraint  for the simple $N_{\rm eff}$ extension, having as much statistical power as the combined BAO dataset.  
By comparing our Fig. \ref{fig:evidencemulti}  with Fig. 6 of Ref.~\cite{paper1} we can appreciate at a glance that, while with pre-{\it Planck} data the Evidence was inconclusive, it is now substantially in favor of $\Lambda$CDM, except when the discrepant $H_0$ is included.

Also particularly interesting are the Evidence ratios for adding massive neutrinos to $\Lambda$CDM, as observations of neutrino oscillations imply neutrinos are not massless. The question that the Evidence ratios address is therefore whether the masses are cosmologically relevant, and the answer from current cosmological data is a resounding ``no'', stronger even than the results from WMAP~\cite{paper1}.  This can be seen by comparing our Fig. \ref{fig:evidencemulti} with its pre-{\it Planck} equivalent: Fig. 6 in Ref.~\cite{paper1}. For all pre-{\it Planck}  dataset combinations the Evidence was indecisive, only becoming substantially in favor of $\Lambda$CDM when the $H_0$ measurement  and the CMB lensing signal from SPT were included. Now the Evidence is strongly in favor of $\Lambda$CDM for all combinations in which an extra dataset is added to the CMB.

Let us note again that Bayesian model comparison rewards {\em predictive} rather than {\em simple} models. If $\Lambda$CDM+$M_\nu$ predicted power spectra that fit the data as well as those predicted by $\Lambda$CDM then the Evidence values for the two models would be equal. This is not the case here: $\Lambda$CDM+$M_\nu$ is disfavored because, on average, summed neutrino masses in the specified prior range provide a worse fit to the data than $\Lambda$CDM. 
Cosmological data do not yet have sufficient precision such that the likelihood for $M_\nu>0$ is significantly  and sufficiently better than that for $M_\nu=0$ to tilt the verdict of the  Evidence ratio.

It is interesting to note that adding massive neutrinos to $\Lambda$CDM is particularly disfavored by the Evidence for the data combination(s) that involve $H_0$. This is  again due to the  mild tension  between the CMB-extrapolated $H_0$ value and the local measurement of the same quantity, which is worsened by increasing the neutrino mass.

When even more complex models are considered, with two extra parameters rather than one (i.e.,  adding $N_{\rm eff}$ and $M_{\nu}$, or $N_{\rm eff}$ and $m_{\nu {\rm sterile}}^{\rm eff}$) the preference of the Evidence ratio for the standard $\Lambda$CDM model increases. This is another consequence of the fact that the standard $\Lambda$CDM model is a good fit to the data, and that more complex models worsen the fit, on average, over the prior ranges spanned by the extra parameters.

%%%%%%%%%%%%%%%%%%%%%%%%%%%%%%%%%%%%%%%%%%%%%%%%%%

\subsection{Profile likelihood ratios and comparison with Evidence}

We compute the PLR for the $\Lambda$CDM+$\neff$ model as outlined in Sec.~\ref{sec:PLR}, using the same dataset combinations as for the Evidence calculations (see Table~\ref{tab:fullresultevidence}). As with the Evidence ratios, the estimates of the PLRs from MCMCs are affected by sampling and binning errors. To estimate the sampling error on the PLR, we first find the difference between the maximum $\Lambda$CDM likelihood and the maximum $\Lambda$CDM+$\neff$ likelihood in a small bin of width 0.01 centered on $\neff = 3.046$.\footnote{Where $\Lambda$CDM chains are available. Where they are not available, we substitute a typical value for this error, 0.7.} By weighting this difference by the square root of the ratio of the number of samples in each PLR bin to the number of samples in the small $\neff \simeq 3.046$ bin, we obtain the errors on the binned PLR. We ensure that the PLR is robust to reasonable changes in the bin size.

\begin{figure}[tb]
\centering
\includegraphics[width=7.5cm]{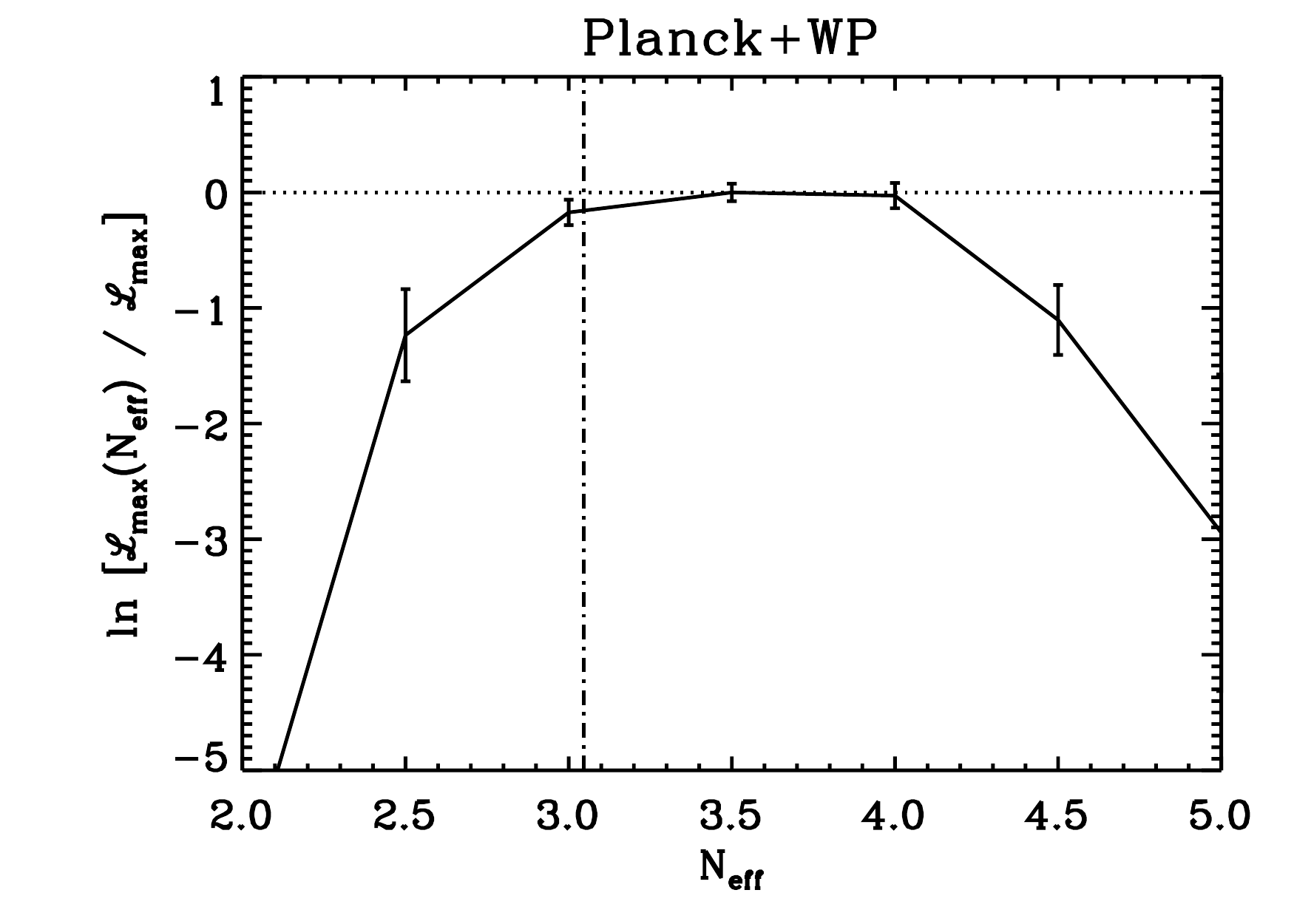} \includegraphics[width=7.5cm]{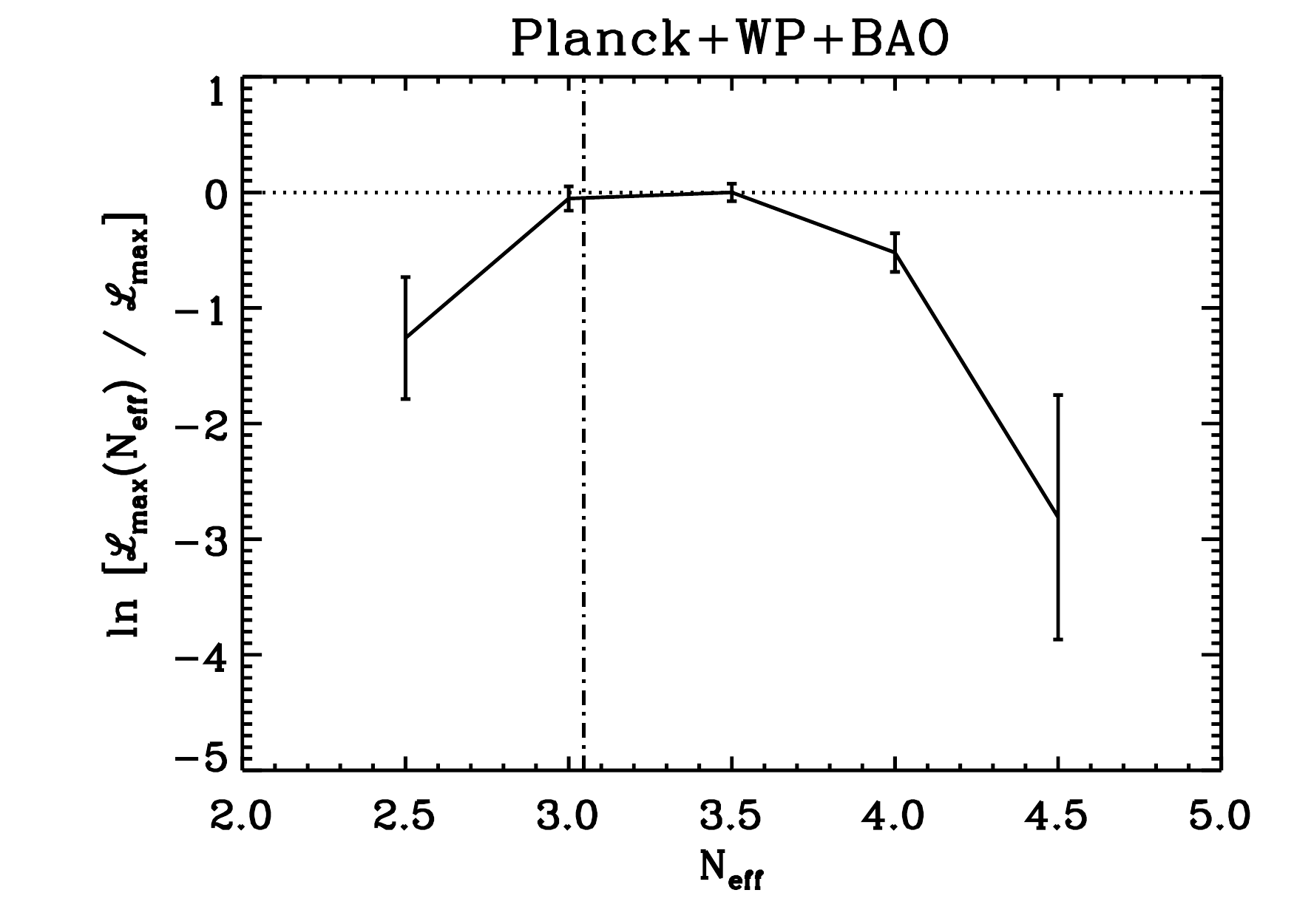} \includegraphics[width=7.5cm]{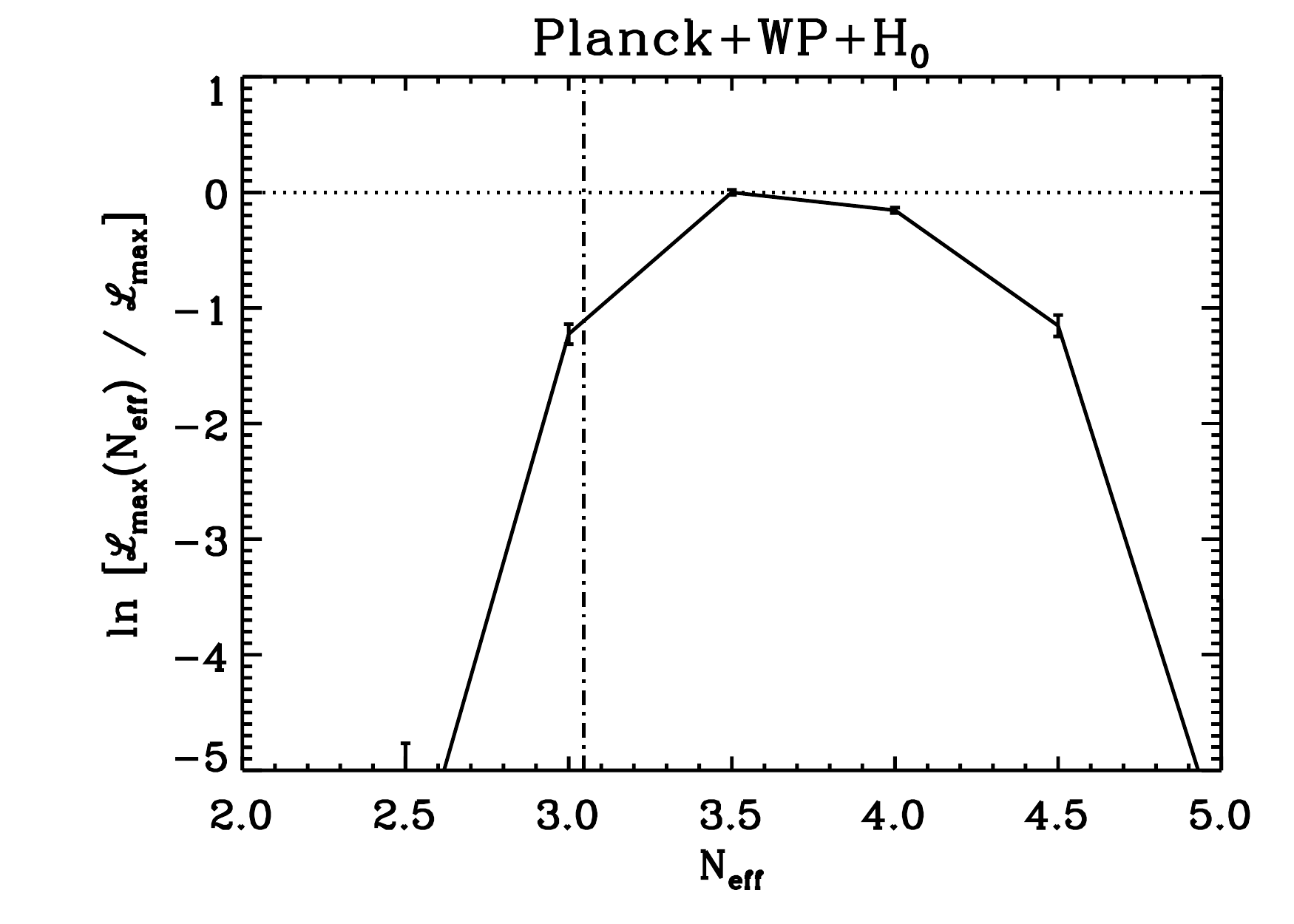}\includegraphics[width=7.5cm]{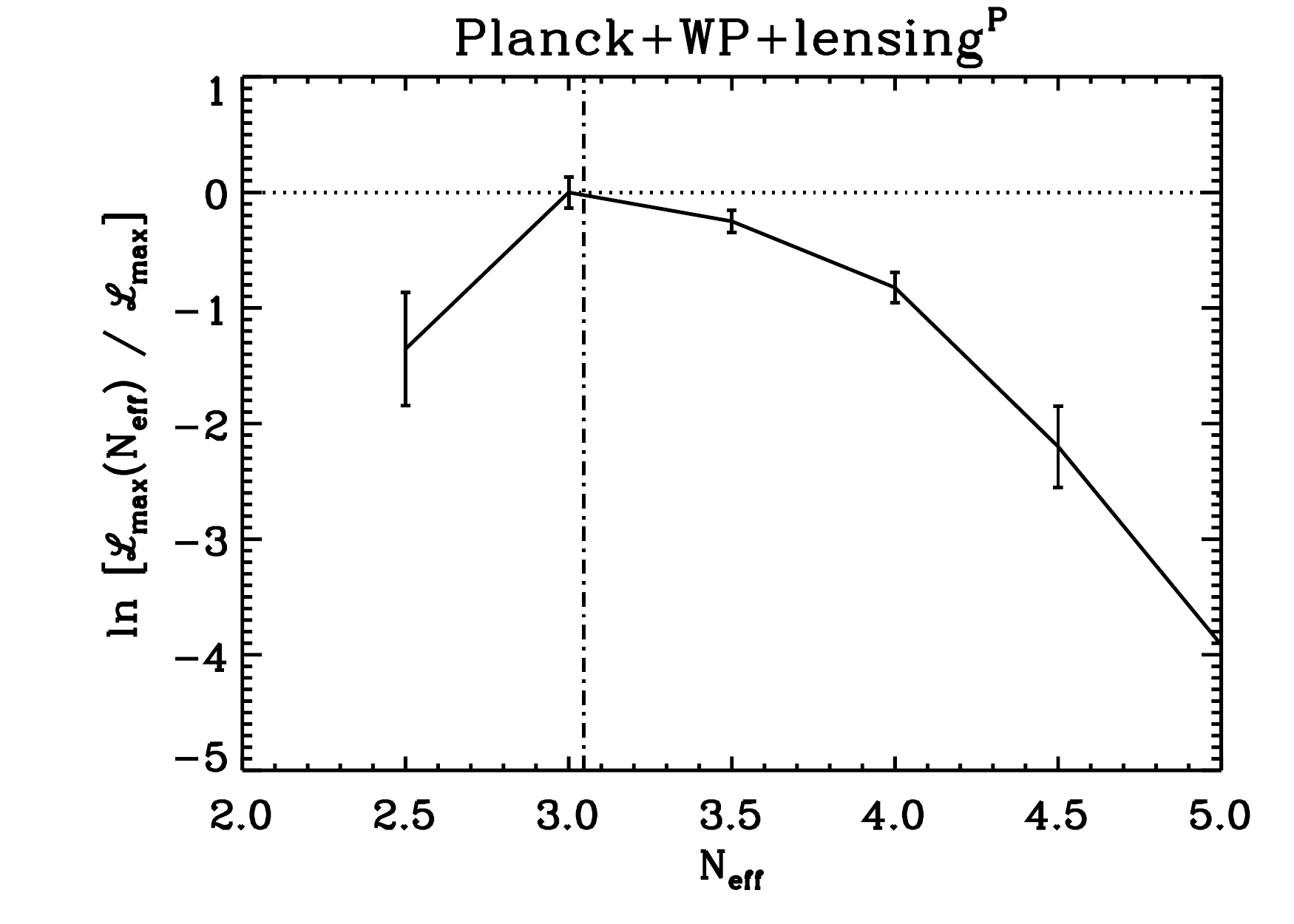}
\caption{Profile likelihood ratio for the {\it Planck} ``Gold'' set  alone and combined with BAO, $H_0$ and  lensing data. The error bars  for the Gold and Gold+lensing sets are computed using the corresponding $\Lambda$CDM chain as described in the main text; errors must be estimated for the other two plots as $\Lambda$CDM chains are not available.  The  dot-dashed vertical lines show the standard $N_{\rm eff}$ value of 3.046.}
\label{fig:PLR}
\end{figure}

The resulting PLR curves are plotted in Fig.~\ref{fig:PLR} for a subset of the data combinations considered. The increase in discriminatory power represented by the {\it Planck} data is clear to see through comparison of this figure with its pre-{\it Planck} equivalent (Fig. 7 in Ref.~\cite{paper1}). We find that the PLR at the standard value of $\neff$ is within errors of its maximum for all dataset combinations apart from those involving the ``discrepant'' $H_0$ measurement, and even in this case the preference for non-standard $\neff$ is less than 2 ``effective" standard deviations. This corroborates, in a prior-independent manner, the inference from the Evidence that the simpler $\Lambda$CDM model is preferred compared to a model with non-standard neutrino properties.

The PLR results are consistent with the Evidence findings: the maximum-likelihood parameters within the extended model are not significantly different from the $\Lambda$CDM values. 
The PLR also shows that, in agreement with the analysis of the posterior~\cite{planckparameterspaper}, values of $N_{\rm eff}\sim 4$ cannot be excluded, but  $N_{\rm eff}\sim 5$  is disfavored.  This is {\it not} in conflict with the Evidence findings: if we had a compelling reason to add one more (effective) neutrino to our base model (i.e., to fix  $N_{\rm eff}$ to $3.046+1=4.046$), the Evidence would still mildly favor this simpler model over a model in which $N_{\rm eff}$ is a free parameter (albeit with $\ln (E_1/E_2) \lesssim 2$, and in many cases below $1$).  It is only if the base model had $5$ neutrino species that the Evidence would favor the addition of $N_{\rm eff}$ as a parameter of the model. In this case, many data combinations (in fact, all combinations involving {\it Planck}+WP), would indicate {\it strong} and even {\it highly significant} Evidence for the $\Lambda$CDM$+N_{\rm eff}$ model.\footnote{For the  $\Lambda$CDM+$Y_P$+$N_{\rm eff}$ model the evidence is slightly weaker; however, for the  {\it Planck}+WP+BAO,  {\it Planck}+WP+lensing,  and {\it Planck}+WP+highL+BAO data combinations, the Evidence for a departure of $N_{\rm eff} $ from the fiducial value would still be {\it strong} and even {\it highly significant}. Similarly, we would have to assume a fixed $M_{\nu}=0.5$ eV in the fiducial model for the Evidence to indicate (at a {\it strong} or  {\it highly significant} level) the need for $M_{\nu}$ as an extra parameter for at least some of the data combinations ({\it Planck}+WP+BAO/$H_0$, {\it Planck}+WP+lensing+BAO/$H_0$ and {\it Planck}+WP+highL+BAO/$H_0$).}

%%%%%%%%%%%%%%%%%%%%%%%%%%%%%%%%%%%%%%%%%%%%%%%%%%

\subsection{Exploring degeneracies with other interesting parameters: the case of $n_s$}

It is well known that the parameters describing neutrino properties are, at least in part, degenerate with other cosmological parameters. The effect of the degeneracy with $Y_P$, e.g., is clearly described in Ref.~\cite{Hou:2011ec}, but the optical depth to reionization, $\tau$, and, more importantly, the spectral slope of the power spectrum, $n_s$, also exhibit degeneracies with $\neff$. The $n_s$ case is particularly interesting.
 With pre-{\it Planck} data (see e.g., Ref.~\cite{Li:2013gka} and references therein)  a scale invariant power spectrum  ($n_s=1$) is consistent at the 95\% confidence level when $N_{\rm eff}$ is allowed  to vary as a parameter of the model.  Within the $\Lambda$CDM model, {\it Planck} data rule out exact scale invariance  at more than $5 \sigma$, indicating strong support for the inflationary paradigm. However, as discussed in Ref.~\cite{Planckinflationpaper}, a model with $n_s=1$ and $N_{\rm eff}\sim 4$ fits the post-{\it Planck} CMB data almost as well as $\Lambda$CDM, although $\Lambda$CDM provides a much better fit when BAO data are also considered. As extensively discussed in e.g., Refs.~\cite{Hou:2011ec, planckparameterspaper}, the BAO measurements (by constraining the expansion history) are sensitive to the matter density, while the CMB geometric degeneracy implies higher cold dark matter density for higher $\neff$.

These conclusions are based on local values of the likelihood: the maximum likelihood for a model with fixed $n_s = 1$ and varying $\neff$ (denoted HZ+$\neff$) is compared to the maximum likelihood for $\Lambda$CDM (with varying $n_s$ and fixed $\neff = 3.046$). The scatter in such measurements is typically fairly large, and the comparison of maximum likelihoods does not provide a self-consistent model-selection criterion; thus it is interesting to determine whether the Evidence can discriminate between the two models. As the HZ+$\neff$ and $\Lambda$CDM models are both nested within $\Lambda$CDM+$\neff$ (at $n_s = 1$ and $\neff = 3.046$, respectively), it is possible to form the desired Evidence ratio $E_{\Lambda {\rm CDM}} / E_{{\rm HZ+}\neff}$ through two applications of the SDDR, determining the Evidence ratios with respect to $\Lambda$CDM+$\neff$ as follows:

\begin{eqnarray}
\frac{E_{\Lambda CDM}}{E_{HZ+N_{\rm eff}}} & = & \frac{ \prob(\data|\Lambda {\rm CDM}) }{ \prob(\data|{\rm HZ+}\neff) } \nonumber\\
& = & \left. \frac{ \prob(\neff | \data, \Lambda {\rm CDM+}\neff) }{ \prob(\neff | \Lambda {\rm CDM+}\neff) } \right|_{\neff = 3.046} \left. \frac{ \prob(n_s | \Lambda {\rm CDM+}\neff) }{ \prob(n_s | \data, \Lambda {\rm CDM+}\neff) } \right|_{n_s = 1}.
\end{eqnarray}

The results of applying the SDDR to the {\it Planck} chains are presented in Table~\ref{tab:HZevidence}. The Evidence ratios support the findings of the {\it Planck} Collaboration's maximum-likelihood analysis. For the Gold dataset, the Evidence ratio is fairly agnostic: there is at most a slight preference for $\Lambda$CDM. When BAO or lensing data are added, however, the Evidence is much more strongly in favor of $\Lambda$CDM.  As discussed above, BAO and lensing constrain the matter density, breaking the parameter degeneracy for CMB data between $\neff$ and cold dark matter.   Interestingly, when HST data are considered there is substantial Evidence in favor of the HZ+$\neff$ model, as $\neff$, $n_s$ and $H_0$ are all positively correlated, and high values of $\neff$ and $n_s$ can therefore resolve the tension between the CMB prediction and local measurements of $H_0$.

 \begin{table}[tb]
\centering
 \begin{tabular}{|c|cccccc|}
 \hline
& \multicolumn{6}{c|}{evidence ratio: $\ln (E_{\Lambda {\rm CDM}} / E_{{\rm HZ+}\neff})$} \\
 \hline
data & base&+BAO&+$H_0$&+SNLS&+Union2 &+lensing\\
 \hline
Planck+WP & 0.95 & 3.23 & -1.86 & -0.11 & 0.66 & 1.70 P\\
  \hline
\end{tabular}
 \caption{Evidence results for the comparison of $\Lambda$CDM to a model with a scale-invariant power spectrum and non-standard effective number of neutrino species, HZ+$\neff$. As previously, datasets marked ``P" indicate that the results have been obtained by importance sampling (post-processing) the MCMCs; the extra datasets are always added individually to the base combination. Positive numbers mean that the simpler model is preferred. The errors on the evidence ratios are in most cases about $\pm 0.02$, with maximum values of $\pm 0.1$ for thinned, post-processed chains.}
 \label{tab:HZevidence}
\end{table}

%%%%%%%%%%%%%%%%%%%%%%%%%%%%%%%%%%%%%%%%%%%%%%%%%%

\section{Conclusions and future avenues}
\label{sec:futureconclusions}

We have used Bayesian model comparison to determine whether extensions to Standard-Model neutrino physics in the form of additional effective numbers of neutrinos and/or massive neutrinos are merited by the latest cosmological data. The official {\it Planck} analysis and subsequent analyses using {\it Planck} data have concentrated on parameter estimation, and have therefore not addressed the model comparison issue considered in this work. We find that the  state-of-the-art cosmological data do not show (Bayesian) Evidence for deviations from the standard $\Lambda$CDM cosmological model (which has three massless neutrino families), findings that are corroborated by a prior-independent analysis based on the PLR. This is particularly interesting for two reasons.

Firstly, perhaps the strongest evidence the {\it Planck} data have provided for inflation is ruling out exact scale-invariance at $\sim 5 \sigma$. As the official {\it Planck} analysis acknowledges, this statement is model dependent: the strongest constraint derives from a comparison of vanilla $\Lambda$CDM with a model with a Harrison Zel'dovich power spectrum, but the constraint is significantly relaxed when the scale-invariant model contains additional effective neutrino species~\cite{planckparameterspaper}. We have bolstered the {\it Planck} team's maximum-likelihood-ratio investigation by determining whether the Bayesian Evidence prefers a model with a Harrison-Zel'dovich power spectrum and freely-varying $\neff$ over standard $\Lambda$CDM. Using only primary CMB data, the Evidence indicates that there is at most a slight preference for $\Lambda$CDM, but when BAO or CMB lensing data are added, this preference increases significantly. While there is considerable variation in the results for other dataset combinations, likely stemming from tensions with the Planck data, it appears that the clear inflationary prediction of near, but not exact, scale-invariance is confirmed by cosmological data even when additional relativistic degrees of freedom are allowed.

Secondly, when considering one-parameter  extensions of the $\Lambda$CDM model described by the addition of an effective number of neutrino species $N_{\rm eff}$, the pre-{\it Planck} cosmological data have been often interpreted as supporting the case for  dark radiation (i.e., deviations from the standard number of $N_{\rm eff}=3.046$), in turn interpreted as evidence for the existence of one or more sterile neutrinos. A Bayesian model-selection analysis did not find significant evidence in favor of extended neutrino physics~\cite{paper1}; nor, however, was it decisively in favor of the Standard Model. Post-{\it Planck}, the data leave less freedom to the sterile-neutrino interpretation, but Bayesian parameter estimates still leave room for dark radiation.  In the present work we find that the Evidence for the  simpler $\Lambda$CDM model has  \emph{increased} with the improvement in observational data represented by {\it Planck}, both alone and in combination with other datasets. The Evidence for the simpler model is ``substantial" in the Jeffreys' scale for many dataset combinations. Even with the discrepant local Hubble constant measurement, there is no Evidence for deviations from the standard $N_{\rm eff}$. The Evidence findings are corroborated by the {\it prior-independent} PLR results. The {\it Planck} Collaboration used the comparison of maximum likelihoods -- a similar approach to the PLR -- to determine any preference for non-standard $\neff$ in selected dataset combinations. Our PLR and Evidence results are in qualitative agreement with the findings of the official {\it Planck} analysis.
 
The next {\it Planck} data release is expected to significantly improve the constraints on $N_{\rm eff}$ due to the additional information contained in the polarization signal.  Changes in $N_{\rm eff}$ leave their specific signature in the polarization as thoroughly explained in Ref. \cite{BashSelk} (see their Figs. 5 and 6 for the effect on the angular power spectrum).  A Fisher-based forecast for the Evidence was presented in Ref.~\cite{KitchingEvidenceFisher}, finding that $\Delta N_{\rm eff}=1$ could be distinguished at a level between ``substantial" and ``strong", even considering a two-parameter extension in which both $N_{\rm eff}$ and $M_{\nu}$ are free parameters. In this work, {\it Planck} Blue Book performance~\cite{PlanckBlueBook} was assumed, and the Evidence was computed under the approximation of Gaussian posteriors~\cite{HeavensEvidence}, but this should not change the conclusions qualitatively. 

In the standard $\Lambda$CDM model neutrinos are massless. This makes the model strictly incorrect, as neutrino oscillations show that neutrinos have a non-zero mass. For this reason, the {\it Planck} team fixed the summed neutrino masses to a value close to the minimum allowed by oscillations in their ``base"  $\Lambda$CDM model.  We have considered whether current cosmological data exhibit any Evidence for this (well motivated)  ``paradigm shift".  We find that the Evidence does not favor the extension of  $\Lambda$CDM to include non-zero neutrino mass: in fact, the Evidence for the standard $\Lambda$CDM model is ``substantial" to ``strong", depending on the dataset combination, and is stronger than it was using pre-{\it Planck} data. 

This  does not conflict with the evidence from neutrino oscillations for non-zero neutrino mass: we consider only cosmological data, which implies that there is no need for an extra parameter, $M_{\nu}$, to describe cosmological data. This only means that $M_{\nu}$ is too small to be cosmologically relevant,
i.e., smaller than the uncertainties of current cosmological measurements. Future data  are expected to have enough statistical power to detect the effect of non-zero neutrino masses with high significance, even if $M_{\nu}$ is close to the minimum mass given by oscillation experiments (see, e.g., Refs.~\cite{Carbonenu, Audren:2012vy, Basse:2013zua, Hamann:2012fe}). However, in the context of current cosmological data, the Bayesian Evidence indicates that the standard $\Lambda$CDM  model does not require any of the  extensions we have considered.

%%%%%%%%%%%%%%%%%%%%%%%%%%%%%%%%%%%%%%%%%%%%%%%%%%

\acknowledgments{LV is supported by  the European Research Council under the European Community's Seventh Framework Programme FP7-IDEAS-Phys.LSS 240117. SMF is supported by STFC  and a grant from the Foundational Questions Institute (FQXi) Fund, a donor-advised fund of the Silicon Valley Community Foundation on the basis of proposal FQXiRFP3-1015 to the Foundational Questions Institute. HVP is supported by STFC, the Leverhulme Trust, and the European Research Council under the European Community's Seventh Framework Programme (FP7/2007-2013) / ERC grant agreement no 306478-CosmicDawn. We acknowledge the use of the Legacy Archive for Microwave Background Data Analysis (LAMBDA), part of the High Energy Astrophysics Science Archive Center (HEASARC). HEASARC/LAMBDA is a service of the Astrophysics Science Division at the NASA Goddard Space Flight Center. This work is based on observations obtained with {\it Planck} (\url{http://www.esa.int/Planck}), an ESA science mission with instruments and contributions directly funded by ESA Member States, NASA, and Canada. The development of {\it Planck} has been supported by: ESA; CNES and CNRS/INSU-IN2P3-INP (France); ASI, CNR, and INAF (Italy); NASA and DoE (USA); STFC and UKSA (UK); CSIC, MICINN and JA (Spain); Tekes, AoF and CSC (Finland); DLR and MPG (Germany); CSA (Canada); DTU Space (Denmark); SER/SSO (Switzerland); RCN (Norway); SFI (Ireland); FCT/MCTES (Portugal); and PRACE (EU). This research was supported in part by the National Science Foundation under Grant No. PHY11-25915.
}

\bibliographystyle{JHEP}
\providecommand{\href}[2]{#2}\begingroup\raggedright\endgroup

%\bibliography{evidence}

\end{document}